\documentclass{iopart}
\usepackage{graphicx}
\usepackage{iopams}
\newcommand{\be}{\begin{equation}}
\newcommand{\ee}{\end{equation}}
\renewcommand{\tr}{{\rm Tr}}
\newcommand{\Wg}{{\rm W}}
\newcommand{\Z}{\mathcal{Z}}
\renewcommand{\S}{\mathcal{S}}

\begin{document}

\title{Semiclassical matrix model for quantum chaotic transport with time-reversal symmetry}
\author{Marcel Novaes}
\address{Instituto de F\'isica, Universidade Federal de Uberl\^andia, Uberl\^andia, MG, 38408-100, Brazil}
\ead{marcel.novaes@gmail.com}

\begin{abstract}We show that the semiclassical approach to chaotic quantum transport in
the presence of time-reversal symmetry can be described by a matrix model, i.e. a matrix
integral whose perturbative expansion satisfies the semiclassical diagrammatic rules for
the calculation of transport statistics. This approach leads very naturally to the
semiclassical derivation of  universal predictions from random matrix theory.
\end{abstract}

\pacs{05.45.Mt,03.65.Sq,73.23.Ad}


\section{Introduction}

We consider wave scattering through a system in which the corresponding ray dynamics is
strongly chaotic. Under a minimal information statistical approach, it is natural to
model the scattering $\S$ matrix as a random matrix, which must be unitary by
conservation principles \cite{blumel,melo,jala}. The quantity $\S_{ij}$ is the scattering
amplitude from channel $j$ to channel $i$. In the presence of time-reversal symmetry,
scattering from $i$ to $j$ is equivalent to scattering from $j$ to $i$ and so $\S$ must
also be symmetric. The ensemble of unitary symmetric complex matrices has a natural
probability measure on it, and is known and the Circular Orthogonal Ensemble (COE) of
random matrix theory (RMT).

RMT predicts successfully several scattering observables, in agreement with numerical
simulations and experimental results, such as average conductance, conductance
fluctuations, average shot-noise and higher counting statistics. This approach has been
reviewed in \cite{beenakker}, and some recent results include
\cite{simm,novaes,savin,vivo,macedo,Anlage}. It is rather flexible, and can be adapted in
order to treat the statistics of time delay \cite{time1,time2,time3}, and to consider the
presence of superconductors \cite{super1,super2}, non-ideal contacts
\cite{contacts1,contacts2,contacts3,contacts4,contacts5}, graphene
\cite{graphene1,graphene2}, etc.

Recovering RMT results from chaotic scattering trajectories has long been a central
problem for the semiclassical approach to quantum mechanics, in which $\S_{ij}$ is
expressed as a sum over paths leading from $j$ to $i$ \cite{semic1,semic2}. In order to
reproduce quantum effects, it is necessary to consider trajectories that are
action-correlated on the scale of $\hbar$. This started to be done perturbatively in
\cite{espalha,lead1} and was shown to give the exact result for the simplest observables
in \cite{prl96sh2006,shot,njp9sm2007}. Attention then turned to more general transport
statistics \cite{jpa41gb2008,njp13gb2011,epl,combinat}, until complete equivalence was
shown between semiclassics and RMT \cite{GregJack0,GregJack} (in the meantime, semiclassics
was able to go beyond RMT, incorporating effects due to finite Ehrenfest time, see e.g. \cite{ehren1,ehren2,ehren3}).

These semiclassical works rely on elaborate manipulations with diagrams and/or
permutations. A more direct demonstration of the RMT-semiclassics equivalence was
presented in \cite{MatrixModel}, based on a matrix model formulation of the semiclassical
approximation. This is a matrix integral whose diagrammatic expansion satisfies exactly
the same diagrammatic rules as the semiclassical calculation of transport observables,
and which turns out to be equivalent to usual RMT.

However, the treatment in \cite{MatrixModel} concerns only systems where time-reversal
symmetry is broken. The purpose of this work is to extend the matrix model approach to
time-reversal symmetric systems, establishing the RMT-semiclassics equivalence in a
direct way for this universality class.

\section{Usual RMT treatment}

Suppose a chaotic cavity, coupled to two ideal leads supporting $N_1$ and $N_2$ open
channels, having a $M-$dimensional $\S$ matrix where $M=(N_1+N_2)$. The energy of the
incoming wave is $E$, and the classical dynamics in the cavity is assumed fully chaotic
at this energy. Under this assumption, RMT assumes the statistical properties of
transport to be independent of $E$.

We shall consider the quantities \be\label{nonlinear}
 P(\vec{i},\vec{j})=\S_{i_1i_2}\S_{i_3i_4}\cdots \S_{i_{2n-1}i_{2n}}\S^*_{j_1j_2}\S^*_{j_3j_4}
 \cdots \S^*_{j_{2n-1}j_{2n}},\ee which can be used to expand any
generic observable (all matrices are taken at the same energy). For instance, the
(dimensionless) conductance and shot-noise are given by \be
g=\sum_{i_1,i_2}\S_{i_1i_2}\S^*_{i_1i_2},\quad
p=g-\sum_{i_1,i_2,i_3,i_4}\S_{i_1i_2}\S^*_{i_3i_2}\S_{i_3i_4}\S^*_{i_1i_4},\ee where the
sums over $i_2,i_4$ ($i_1,i_3$) run over the $N_1$ incoming ($N_2$ outgoing) channels.

According to the random matrix theory approach, for time-reversal invariant chaotic
systems $\S$ is uniformly distributed in the COE$(M)$. The interest then lies in the
average value of (\ref{nonlinear}) over this ensemble. The simplest case is
\be\label{conduc} \langle |\S_{ij}|^2\rangle_{\rm COE(M)}=\frac{1+\delta_{ij}}{M+1},\ee
which leads to the well known prediction $N_1N_2/(M+1)$ for the average conductance. In
general, the average vanishes unless $\vec{j}$ is equal to some permutation of $\vec{i}$.
For instance, \be\label{S4} \langle \S_{12}\S^*_{12}\S_{34}\S^*_{34}\rangle_{\rm COE(M)}
=\frac{M+2}{M(M+1)(M+3)},\ee and \be\label{S4b}  \langle
\S_{12}\S^*_{14}\S_{34}\S^*_{23}\rangle_{\rm COE(M)}=\frac{-1}{M(M+1)(M+3)}.\ee

There may be several possibilities for the permutation relating $\vec{j}$ and $\vec{i}$,
and the general expression for our average contains a sum, \be\label{COE} \left\langle
P(\vec{i},\vec{j})\right\rangle_{\rm COE(M)}= \sum_{\pi \in \mathfrak{S}_{2n}}
\delta_\pi(\vec{i},\vec{j}) \Wg_{M}(\pi),\ee where $\mathfrak{S}_n$ is the permutation
group of $n$ symbols and \be
\delta_\pi(\vec{i},\vec{j})=\prod_{k=1}^{2n}\delta_{i_k,j_{\pi(k)}}.\ee The quantity
$\Wg_{M}(\pi)$, called the Weingarten function of the COE, can be computed from explicit
character-theoretic formulas \cite{mats} or via recurrence relations \cite{BB}.

Since the $\S$-matrix is symmetric, the value of (\ref{COE}) must be invariant under the
action of any permutation that interchanges $j_{2k-1}$ with $j_{2k}$, or any permutation
that interchanges simultaneously $j_{2k-1}$ with $j_{2r-1}$ \emph{and} $j_{2k}$ with
$j_{2r}$. The set of all such permutations is called the hyperoctahedral group
$\mathfrak{H}_n\subset\mathfrak{S}_{2n}$. For $n=2$ it consists of the permutations
$\{1,(12),(34),(12)(34)\}$. In general, $\mathfrak{H}_n$ has $2^nn!$ elements. Thus, if
$\xi\in\mathfrak{H}_n$ then \be\label{inv} \Wg_{M}(\pi\xi)=\Wg_{M}(\pi).\ee

\section{Semiclassical Diagrammatics}

The semiclassical approximation to $\S_{ij}$ requires trajectories starting at channel
$j$ and ending at channel $i$. In the semiclassical expression for (\ref{nonlinear}) we
end up with some trajectories (\emph{direct} ones) going from $i_{2k}$ to $i_{2k-1}$ and
some other trajectories (\emph{partner} ones) going from $j_{2k}$ to $j_{2k-1}$. Consider
now the average value of (\ref{nonlinear}) over a certain energy window, $\langle
P(\vec{i},\vec{j})\rangle_E$, this window being small in the classical scale but large in
the quantum scale; as $\hbar\to 0$, constructive interference is required and the result
is determined by correlations: partner trajectories must have almost the same collective
action as direct ones.

The theory of correlated chaotic trajectories has been discussed in detail in a number of
papers \cite{prl96sh2006,shot,njp9sm2007,njp13gb2011,GregJack}. Trajectories from
correlated sets may differ only in small regions (called \emph{encounters}) in which the
direct ones run nearly parallel or anti-parallel, while the partner ones have crossings.
This ensures they have almost the same collective action. In particular, this implies
that $\vec{i}$ and $\vec{j}$ must be equal up to a permutation, a condition already met
in the RMT treatment.

These trajectory multiplets are usually represented by diagrams, in which the complicated
pieces of chaotic trajectories in-between encounters are depicted as simple links.
Calculation of any given $\langle P(\vec{i},\vec{j})\rangle_E$ requires constructing all
possible contributing diagrams. Most importantly, a diagrammatic rule has been devised
for the value of any diagram: it is $(-1)^VM^{V-L}$, where $V$ is the number of
encounters and $L$ is the number of links.

\begin{figure}
\includegraphics[scale=0.65,clip]{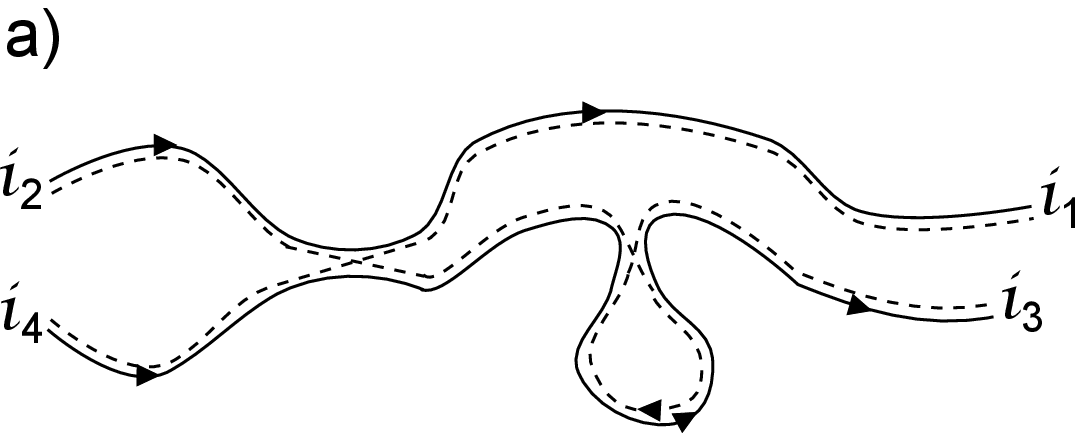}\quad \quad\includegraphics[scale=0.65,clip]{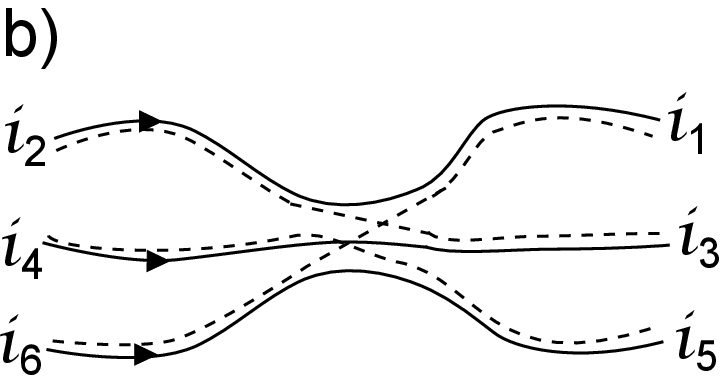}
\caption{Two examples of semiclassical diagrams required in the calculation of transport
observables. Direct trajectories (solid lines) are correlated with partner trajectories
(dashed lines), by means of $2$-encounters (left) and a $3$-encounter (right). This is
very simplified: actual trajectories are long and chaotic.}
\end{figure}

\begin{figure}
\includegraphics[scale=0.65,clip]{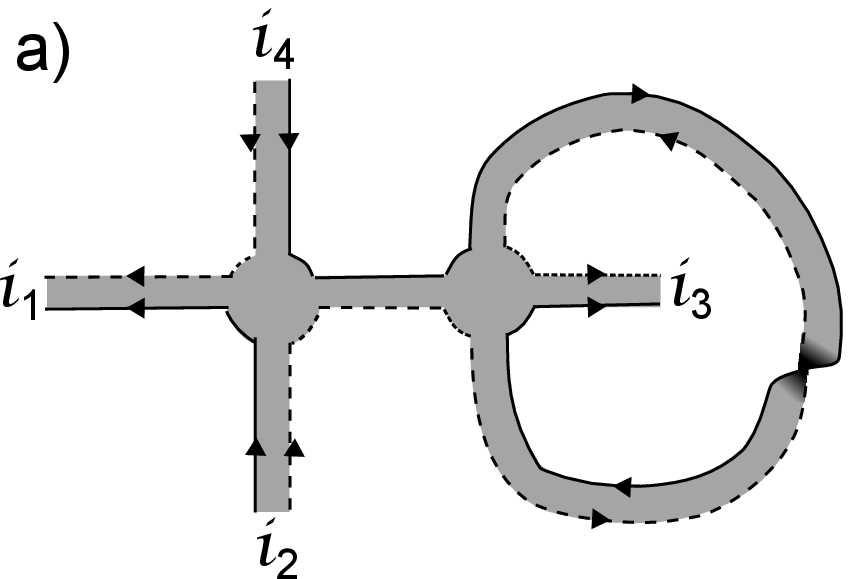}\quad \quad \quad \includegraphics[scale=0.65,clip]{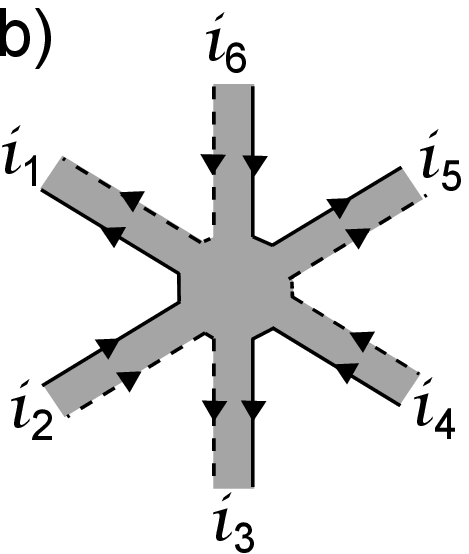}
\caption{A different representation of the semiclassical diagrams of Figure 1, in terms
of ribbon graphs. Encounters become vertices and trajectories become edges of ribbons.
Notice how one of the ribbons must be twisted in the first case.}
\end{figure}

We show two examples in Figure 1. Panel a) shows a contribution to
$\S_{i_1i_2}\S^*_{i_3i_2}\S_{i_3i_4}\S^*_{i_1i_4}$, containing two $2$-encounters. One of
the encounters involves two different trajectories, while the other one is of a
trajectory with itself. Notice how the direct and partner trajectories run in opposite
senses in one of the regions. This contribution is only possible because we are assuming
the system has time-reversal symmetry. Its value is $1/M^4$. Panel b) shows a
contribution to
$\S_{i_1i_2}\S^*_{i_3i_2}\S_{i_3i_4}\S^*_{i_5i_4}\S_{i_5i_6}\S^*_{i_1i_6}$, containing a
single $3$-encounter. This diagram does not require time-reversal symmetry, and its value
is $-1/M^5$.

A different diagrammatic representation, more convenient, of correlated trajectories uses
ribbon graphs \cite{GregJack}. In this case we turn every encounter into a vertex, and
trajectories are depicted as edges of ribbons. It is sometimes necessary to perform
twists on some of the ribbons. The diagrams from Figure 1a,b are represented as ribbon
graphs in Figure 2a,b.

\section{Matrix model for the semiclassical approach}

\subsection{Wick's rule}

Suppose the matrix integral \be \langle f(S)\rangle_S=\frac{1}{\mathcal{Z}}\int dS
e^{-\frac{M}{2}\tr SS^T}f(S),\ee where $M$ is a parameter and \be\mathcal{Z}=\int dS
e^{-\frac{M}{2}\tr SS^T}\ee is a normalization factor. This integral runs over general
$N$-dimensional real matrices, without constraints (this is known as the real Ginibre
ensemble \cite{ginibre}). Notice that the dimension $N$ is not related to channel
numbers; we continue with $M=N_1+N_2$.

Gaussian integrals can be performed exactly, leading to \be\label{cov} \langle
S_{ab}S_{cd}\rangle_S =\frac{\delta_{ac}\delta_{bd}}{M}.\ee When there are many matrix
elements being integrated (the number must be even, otherwise the result vanishes), we
may use the well known Wick's rule, \be\label{wick1}\left\langle\prod_{k=1}^{2n}
S_{a_kb_k}\right\rangle_S =\sum_{\sigma\in \mathfrak{M}_n}\prod_{k=1}^{n}\langle
S_{a_{\sigma(2k-1)}b_{\sigma(2k-1)}}S_{a_{\sigma(2k)}b_{\sigma(2k)}}\rangle_S\ee where
the sum is over all possible matchings among $2n$ elements. For example, at $n=2$ there
are three possible matchings, \begin{eqnarray} &\langle
S_{a_1b_1}S_{a_2b_2}S_{a_3b_3}S_{a_4b_4}\rangle=\langle
S_{a_1b_1}S_{a_2b_2}\rangle\langle S_{a_3b_3}S_{a_4b_4}\rangle\nonumber\\&+\langle
S_{a_1b_1}S_{a_3b_3}\rangle\langle S_{a_2b_2}S_{a_4b_4}\rangle+\langle
S_{a_1b_1}S_{a_4b_4}\rangle\langle S_{a_2b_2}S_{a_3b_3}\rangle.\end{eqnarray}

Elements of $\mathfrak{M}_n$ can be represented by permutations acting on the trivial
matching $\{\{1,2\},\{3,4\},...\}$. In this sense, the above matchings correspond to the
identity permutation, to the transposition $(23)$ and to the cycle $(243)$. A permutation
$\sigma$ represents a matching if and only if it satisfies $\sigma(2i-1)<\sigma(2i)$ and
$\sigma(2i-1)<\sigma(2i+1)$. Any element of $\mathfrak{S}_{2n}$ can be decomposed
uniquely as the product of a member of $\mathfrak{M}_n$ and a member of $\mathfrak{H}_n$,
i.e. the set $\mathfrak{M}_n$ can be seen as the coset
$\mathfrak{S}_{2n}/\mathfrak{H}_n$.

\subsection{Diagrammatics}

Introduce the following diagrammatical representation to the calculation of
(\ref{wick1}). Each matrix element $S_{ab}$ is represented by a ribbon, having one edge
associated with $a$ and depicted with a solid line, and the other edge associated with
$b$ and depicted with a dashed line. Wick's rule then tells us to draw all possible
connections among these lines, and associate to each connection a factor $1/M$. When two
ribbons are connected, their solid edges merge and must have the same label, and likewise
for the dashed edges.

For example, consider the average value \be\label{example} \fl\left\langle {\rm
Tr}(SS^T)^2S_{a_1b_1}S_{a_2b_2}\right\rangle=\sum_{a_3a_4b_3b_4=1}^{N}\left\langle
S_{a_1b_1}S_{a_2b_2}S_{a_3b_3}S_{a_4b_3}S_{a_4b_4}S_{a_3b_4}\right\rangle.\ee The
presence of the trace requires repeated indices among the matrix elements. This is taken
into account by arranging them around a vertex. We show in Figure 3 the ribbons
associated with (\ref{example}), and the possible connections in the application of
Wick's rule. There are six distinct topologies, each appearing with some multiplicity.
The calculation results in the value \be
\frac{-1}{M^3}(2N^3+N^2+8N+4)\delta_{a_1a_2}\delta_{b_1b_2}.\ee The power of $N$ counts
the number of indices being freely summed over, associated with closed lines in the
diagrams.

\begin{figure}
\includegraphics[scale=0.6,clip]{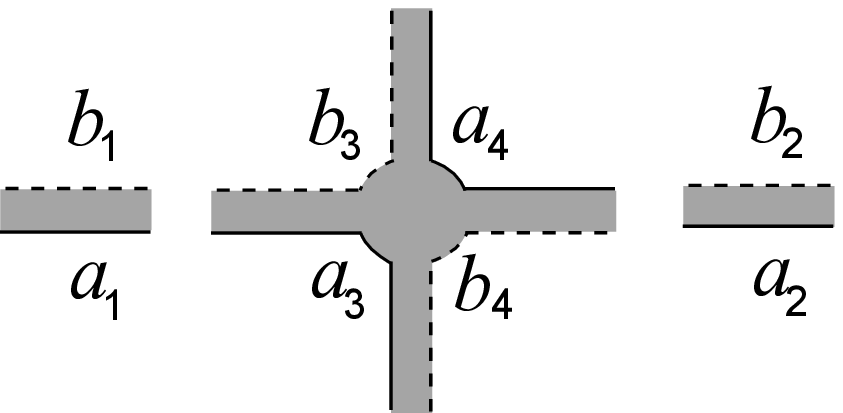}\quad \includegraphics[scale=0.7,clip]{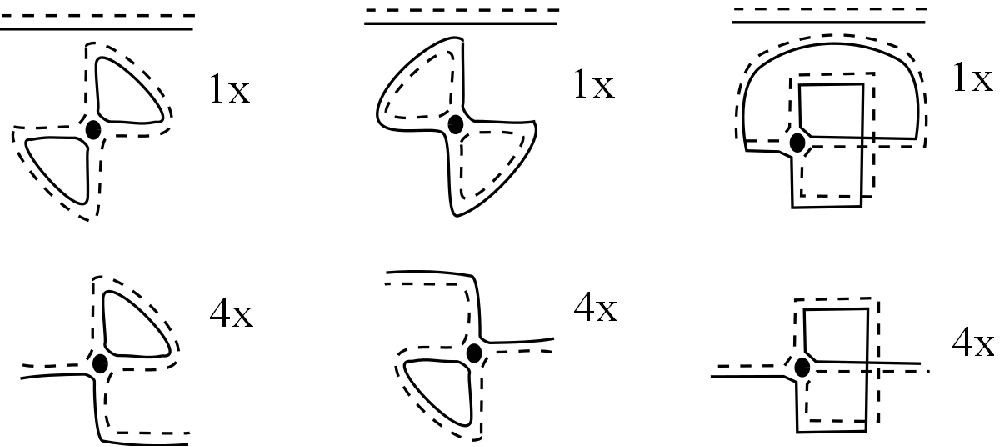}
\caption{Diagrammatical representation of the average value of ${\rm
Tr}(SS^T)^2S_{a_1b_1}S_{a_2b_2}$, and the possible connections between lines, according
to Wick's rule. In the upper line a direct connection is made between $S_{a_1b_1}$ and
$S_{a_2b_2}$. There are in total 15 connections, and we show how many times each topology
appears, due to trivial symmetries.}
\end{figure}

This kind of diagrammatics has been extensively discussed in the literature. The most
common situation is when complex hermitian matrices are used \cite{diagram1}, in which
case no twists are required in the ribbons (they are orientable) and no distinction is
made between the edges. Complex non-hermitian matrices \cite{diagram2} maintain
orientability but allow different edges. For real symmetric matrices the ribbons are not
necessarily orientable \cite{diagram3,diagram4} and edges are indistinguishable. Our case
(general real matrices) does not require orientability, and has distinguished edges.

\subsection{Introducing channels}

We are using the labels $a$ and $b$ to identify the `trajectories' of our matrix model.
We must still introduce the channel labels. To that end, let $Q=1_M\oplus 0_{N-M}$ be an
orthogonal projector, having the $M-$dimensional identity in its upper-left corner and
all other entries equal to zero. Let us also define the matrix \be R=XQSQX^\dag,\ee where
$X$ is some complex $M\times M$ matrix. This leads to \be \prod_{k=1}^{2n}R_{i_ki_k} =
\prod_{k=1}^{2n}\sum_{a_k,b_k=1}^M X_{i_ka_k}S_{a_kb_k}X^*_{i_kb_k}.\ee

The indices $i_k$ represent the channels. The idea is that, upon taking the average over
the $S$ matrices, the $a$-labels will produce the direct trajectories connecting the
channels, while the $b$-labels will produce the partner trajectories, which may connect
the channels in a different way but will necessarily be correlated with the direct ones.

For future reference, let us also define the matrix $Z=XX^T$. Notice that this is a
complex symmetric matrix. The elements of $Z$ will be related to direct trajectories, and
the elements of $Z^*$ will be related to the partner trajectories.

Before going into further details of this method, let us show it in action for the
simplest example.

\subsection{The conductance}

Conductance requires the average value of $\S_{i_1i_2}\S^*_{i_1i_2}$. We start by writing
\be\fl \left\langle
R_{i_1i_1}R_{i_2i_2}\right\rangle_{S}=\sum_{a_1a_2b_1b_2=1}^MX_{i_1a_1}X_{i_2a_2}X^*_{i_1b_1}X^*_{i_2b_2}\left\langle
S_{a_1b_1}S_{a_2b_2}\right\rangle_{S}=\frac{1}{M}Z_{i_1i_2}Z^*_{i_1i_2}.\ee The
coefficient $1/M$ is related to the leading order approximation to Eq.(\ref{conduc}).

Next, we can produce diagrams with encounters by including traces of powers of $SS^T$.
The simplest such case is \be\label{g2} \left\langle \frac{-M}{4}{\rm
Tr}(SS^T)^2R_{i_1i_1}R_{i_2i_2}\right\rangle_{S}=\frac{-1}{4M^2}(4+8N+N^2+2N^3)Z_{i_1i_2}Z^*_{i_1i_2}.\ee
The diagrammatics of this average is the same as the one shown in Figure 3. We have
multiplied by $-M$ because we already know that each encounter must be accompanied by
such factor in the semiclassical diagrammatics. We have divided by $4$ because of the
rotation symmetry of the $2$-encounter.

As we have seen, closed lines produce powers of $N$ (the $S$-matrices inside the trace
are not truncated). In a semiclassical interpretation, such closed lines would represent
periodic orbits, forever trapped inside the system. The true semiclassical diagrams
contain only scattering trajectories, and no periodic orbits. We can get rid of such
orbits by the trick of letting $N\to 0$. If we perform this trick, we get
$Z_{i_1i_2}Z^*_{i_1i_2}$ with a coefficient of $-1/M^2$, which is the second order
approximation to Eq.(\ref{conduc}).

A triple encounter corresponds to \be\label{g3a} \lim_{N\to 0}\left\langle
\frac{-M}{6}{\rm
Tr}(SS^T)^3R_{i_1i_1}R_{i_2i_2}\right\rangle_{S}=\frac{-1}{M^3}Z_{i_1i_2}Z^*_{i_1i_2},\ee
where we have discounted a $6$-fold rotation symmetry. The situation with two single
encounters correspond to \be\label{g3b} \lim_{N\to 0}\left\langle
\frac{1}{2}\left[\frac{-M}{4}{\rm
Tr}(SS^T)^2\right]^2R_{i_1i_1}R_{i_2i_2}\right\rangle_{S}=\frac{2}{M^3}Z_{i_1i_2}Z^*_{i_1i_2}.\ee
Here there is an extra denominator of $2$ to account for the exchange symmetry between
the encounters. The values of (\ref{g3a}) and (\ref{g3b}) together produce a coefficient
to $Z_{i_1i_2}Z^*_{i_1i_2}$ which is $1/M^3$, related precisely to the third order
approximation to Eq.(\ref{conduc}).

In order to produce all possible encounters, we must include all possible traces, each
one of them multiplied by $-M$, as required by the semiclassical diagrammatic rules. We
must also discount an overcounting of $2q$ from rotation symmetry around vertices of
valence $q$, and take into account the exchange symmetry between encounters of the same
valence. Fortunately, all this is automatically implemented by means of an exponential
function.

Therefore, in analogy to \cite{MatrixModel}, our semiclassical matrix model for the
average value of $\S_{i_1i_2}\S^*_{i_1i_2}$ is to compute \be\lim_{N\to 0}\left\langle
e^{-M\sum_{q\ge 2} \frac{1}{2q}\tr[(SS^T)^q]} R_{i_1i_1}R_{i_2i_2}\right\rangle_{S},\ee
and to extract the coefficient of $Z_{i_1i_2}Z^*_{i_1i_2}$. In order to compute the
conductance, we just sum over $i_1,i_2$.

\subsection{Higher Moments}

To treat conductance, which involves two channels, we considered the quantity
$R_{i_1i_1}R_{i_2i_2}$. In order to be able to obtain a semiclassical matrix model for
higher transport moments, involving $n$ channels, we must consider more general
quantities of the form $ \prod_{k=1}^{2n}R_{i_ki_k}$.

As an example, let us look at the diagrammatics of this average value for $n=2$. Using
Wick's rule, it is easy to see that this is given by \begin{eqnarray}\label{R4}
\left\langle
R_{i_1i_1}R_{i_2i_2}R_{i_3i_3}R_{i_4i_4}\right\rangle_{S}=\frac{1}{M^2}\left[Z_{i_1i_2}Z^*_{i_1i_2}
Z_{i_3i_4}Z^*_{i_3i_4}\right.\nonumber\\+\left.Z_{i_1i_3}Z^*_{i_1i_3}Z_{i_2i_4}Z^*_{i_2i_4}+
Z_{i_1i_4}Z^*_{i_1i_4}Z_{i_2i_3}Z^*_{i_2i_3}\right].\end{eqnarray} The channel labels
appear in different arrangements on the right hand side, corresponding to the
semiclassical diagrams in Figure 4a. This quantity is thus producing all the leading
order transport diagrams with four channels simultaneously.

\begin{figure}
\includegraphics[scale=0.8,clip]{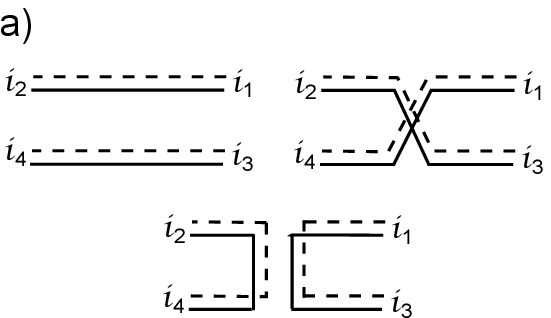}\quad \quad \includegraphics[scale=0.8,clip]{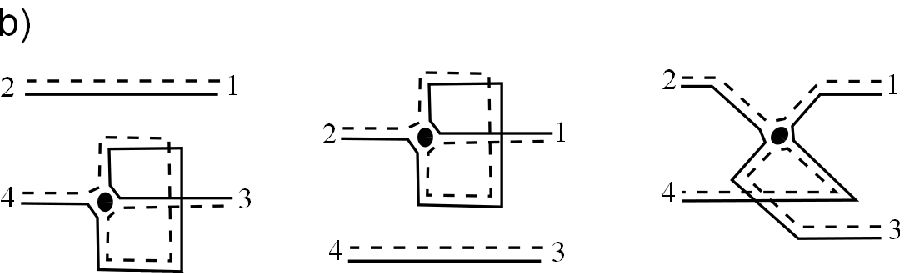}
\caption{a) Diagrams related to the leading order approximation to $\left\langle
R_{i_1i_1}R_{i_2i_2}R_{i_3i_3}R_{i_4i_4}\right\rangle_{S}$. b) The first two diagrams are
responsible for Eq.(\ref{41}), while the last diagram is responsible for Eq.(\ref{41b}).}
\end{figure}

If we wish to know the leading order approximation to the average value of a certain
quantity, say $\S_{i_1i_2}\S^*_{j_1j_2} \S_{i_3i_4}\S^*_{j_3j_4}$, we just extract from
the above result the coefficient of
$\left[Z_{i_1i_2}Z^*_{j_1j_2}Z_{i_3i_4}Z^*_{j_3j_4}\right]$ (when performing this
coefficient extraction procedure, we consider the indices to be independent variables,
i.e. we do not care for possible coincidences among the elements of $\vec{i}$). We shall
use the notation $[x]f$ to denote the coefficient of $x$ in $f$.

For example, looking at Eq.(\ref{R4}) it is clear that \be
\left[Z_{i_1i_2}Z_{i_3i_4}\right]\left\langle
\prod_{k=1}^{4}R_{i_ki_k}\right\rangle_{S}=\frac{1}{M^2}Z^*_{i_1i_2}Z^*_{i_3i_4}.\ee On
the other hand, when we consider
$\left[Z^*_{j_1j_2}Z^*_{j_3j_4}\right]Z^*_{i_1i_2}Z^*_{i_3i_4}$, it follows that
$\vec{j}$ and $\vec{i}$ can only differ by a hyperoctahedral permutation: either we have
the set identities $\{j_1,j_2\}=\{i_1,i_2\},\{j_3,j_4\}=\{i_3,i_4\}$ or the set
identities $\{j_1,j_2\}=\{i_3,i_4\},\{j_3,j_4\}=\{i_1,i_2\}$. That is, we have that \be
\left[Z^*_{j_1j_2}Z^*_{j_3j_4}\right]Z^*_{i_1i_2}Z^*_{i_3i_4}=\sum_{\pi\in\mathfrak{H}_2}
\delta_{\pi}(\vec{i},\vec{j}).\ee In conclusion, our semiclassical prediction for the
leading order approximation to the average value of $\S_{i_1i_2}\S^*_{j_1j_2}
\S_{i_3i_4}\S^*_{j_3j_4}$ is $M^{-2}\sum_{\pi\in\mathfrak{H}_2}
\delta_{\pi}(\vec{i},\vec{j}).$

More concretely, we have, for instance, \be [Z_{12}Z^*_{12}Z_{34}Z^*_{34}]\left\langle
R_{11}R_{22}R_{33}R_{44}\right\rangle_{S}=\frac{1}{M^2},\ee which is indeed the leading
order approximation to Eq.(\ref{S4}), and \be[Z_{12}Z^*_{14}Z_{34}Z^*_{23}]\left\langle
R_{11}R_{22}R_{33}R_{44}\right\rangle_{S}=0,\ee reflecting the fact that there are no
semiclassical diagrams contributing at leading order to the calculation of
Eq.(\ref{S4b}), and that indeed Eq.(\ref{S4b}) is of order $M^{-3}$.

We can produce perturbative corrections, i.e. diagrams with encounters, by including
traces, as we did for the conductance. For instance, a single encounter is introduced as
\be\label{41}\lim_{N\to 0} [Z_{12}Z^*_{12}Z_{34}Z^*_{34}]\left\langle \frac{-M}{4}{\rm
Tr}(SS^T)^2\prod_{k=1}^{4}R_{kk}\right\rangle_{S}= \frac{-2}{M^3},\ee which is the first
correction to Eq.(\ref{S4}), and\be\label{41b} \lim_{N\to
0}[Z_{12}Z^*_{14}Z_{34}Z^*_{23}]\left\langle \frac{-M}{4}{\rm
Tr}(SS^T)^2\prod_{k=1}^{4}R_{kk}\right\rangle_{S}= \frac{-1}{M^3},\ee which is the
leading order contribution to Eq.(\ref{S4b}). The diagrams corresponding to the above
results can be seen in Figure 4b. As already discussed, the limit $N\to 0$ is necessary
to discard diagrams having periodic orbits.

As we saw previously, in order to produce all possible encounters we must introduce the
factor $e^{-M\sum_{q\ge 2} \frac{1}{2q}\tr[(SS^T)^q]}$. For convenience, let us define
\be Z_{\vec{i}}=\prod_{k=1}^nZ_{i_{2k-1}i_{2k}}.\ee Then, our semiclassical matrix model
is \be\label{modela} \left\langle P(\vec{i},\vec{j})\right\rangle_E=\lim_{N\to
0}\left[Z_{\vec{i}}Z^*_{\vec{j}}\right]G(\vec{i}),\ee where \be\label{modelb}
G(\vec{i})=\left\langle e^{-M\sum_{q\ge 2} \frac{1}{2q}\tr[(SS^T)^q]} \prod_{k=1}^{2n}
R_{i_ki_k}\right\rangle_{S}.\ee This matrix integral has, by construction, exactly the
same diagrammatic formulation as the semiclassical approach.

\section{Equivalence with RMT}

We wish to show equivalence between semiclassical theory and random matrix theory, i.e.
between Eqs.(\ref{modela})-(\ref{modelb}) and Eq.(\ref{COE}).

We start by noticing that, because of the identity \be \det(1-A)=e^{{\rm Tr}\log
(1-A)}=e^{-\sum_{q\ge 1}\frac{1}{q}{\rm Tr}(A^q)},\ee we can write \be G(\vec{i})=\int
\frac{dS}{\mathcal{Z}} \det(1-SS^T)^{M/2}\prod_{k=1}^{2n} R_{i_ki_k}.\ee Next, from the
theory of truncated orthogonal matrices it can be established (see Appendix) that
\be\label{limit} \lim_{N\to 0} \int
\frac{dS}{\mathcal{Z}}\det(1-SS^T)^{M/2}\prod_{k=1}^{2n} S_{a_kb_k} =\left\langle
\prod_{k=1}^{2n} O_{a_kb_k}\right\rangle_{\mathcal{O}(M+1)},\ee provided $1\leq
a_k,b_k\leq M$ and where the average on the right-hand-side is over random matrices
uniformly distributed (with respect to Haar measure) in the orthogonal group of
$(M+1)$-dimensional real matrices satisfying $OO^T=1$.

It is known \cite{CS,CM} that this orthogonal group average is given by a double sum over
the set of matchings, \be \left\langle \prod_{k=1}^{2n}
O_{a_kb_k}\right\rangle_{\mathcal{O}(M+1)}=\sum_{\sigma,\tau\in\mathfrak{M}_n}
\Delta_\sigma(a) \Delta_\tau(b)\mathcal{W}_{M+1}(\sigma^{-1}\tau),\ee where the quantity
$\mathcal{W}_{M+1}$ is the Weingarten function of $\mathcal{O}(M+1)$ and \be
\Delta_\sigma(a)=\prod_{k=1}^n \delta_{a_{\sigma(2k-1)},a_{\sigma(2k)}}.\ee Noticing that
\be \sum_{a_1,...a_{2n}=1}^M\Delta_\sigma(a)\prod_{k=1}^{2n}X_{i_ka_k}=\prod_{k=1}^{n}
(XX^T)_{i_{\sigma(2k-1)},i_{\sigma(2k)}}=Z_{\sigma(\vec{i})}\ee leads to \be \lim_{N\to
0}G(\vec{i})=\sum_{\sigma,\tau\in\mathfrak{M}_n}
\mathcal{W}_{M+1}(\sigma^{-1}\tau)Z_{\sigma(\vec{i})} Z^*_{\tau(\vec{i})}. \ee

We now extract the coefficient of $\left[Z_{\vec{i}}Z^*_{\vec{j}}\right]$. The only
matching $\sigma$ which survives this operation is the identity. On the other hand, we
have already seen that
\be\left[Z^*_{\vec{j}}\right]Z^*_{\tau(\vec{i})}=\sum_{\rho\in\mathfrak{H}_n}\delta_\rho(\tau(\vec{i}),\vec{j}).\ee
Thus, we get \be \lim_{N\to
0}\left[Z_{\vec{i}}Z^*_{\vec{j}}\right]G(\vec{i})=\sum_{\tau\in\mathfrak{M}_n}
\sum_{\rho\in\mathfrak{H}_n} \mathcal{W}_{M+1}(\tau)\delta_\rho(\tau(\vec{i}),\vec{j}).
\ee We may use the invariance of the Weingarten function $\mathcal{W}$ under the action
of the hyperoctahedral, $\mathcal{W}_{M+1}(\rho\tau)=\mathcal{W}_{M+1}(\tau)$, to group
both sums into a single sum over the whole permutation group. Defining $\pi=\rho\tau$, we
have \be\left\langle P(\vec{i},\vec{j})\right\rangle_E=\sum_{\pi\in\mathfrak{S}_n}
\mathcal{W}_{M+1}(\pi)\delta_\pi(\vec{i},\vec{j}).\ee

It was shown in \cite{mats} that the Weingarten functions of the orthogonal group and of
the COE are related by the simple identity \be \mathcal{W}_{M+1}(\pi)=\Wg_{M}(\pi).\ee We
are hereby showing that this equality, rooted in the fact that COE($M$) can be seen as
the quotient space $\mathcal{U}(M)/\mathcal{O}(M)$, where $\mathcal{U}(M)$ is the unitary
group, is in fact the key to the equivalence between the semiclassical and RMT approaches
to quantum chaotic transport in the presence of time-reversal symmetry, because it leads
precisely to the fact that \be\left\langle P(\vec{i},\vec{j})\right\rangle_E=\left\langle
P(\vec{i},\vec{j})\right\rangle_{{\rm COE}(M)}.\ee

\section{Conclusions}

We have extended the matrix model approach to semiclassical quantum chaotic transport in
order to treat systems with time-reversal symmetry. The hardest part of the method is
designing a matrix integral with the correct semiclassical diagrammatic rules. Once this
is in place, it leads in a very direct way to the equivalence to random matrix theory.

This approach may open the way to semiclassical calculations that were previously
unavailable, and may even provide results beyond RMT. For example, it may be adapted to
treat problems where the semiclassical trajectories have different energies, as is
necessary in calculations involving time delay (this program has already been carried out
for broken time reversal symmetry in \cite{timemeu}). Another possibility is the
treatment of the proximity gap in Andreev billiards \cite{Andreev}.

Similar ideas might also be applied to closed systems, allowing the calculation of
spectral correlation functions and providing justification for the celebrated
Bohigas-Giannoni-Schmit conjecture \cite{BGS,haakepre} that they are all described by
RMT.

\ack

Gregory Berkolaiko was involved in the initial stages of this project, and interesting
conversations with him are gratefully acknowledged, as well as financial support from
Conselho Nacional de Desenvolvimento Cient\'ifico e Tecnol\'ogico (CNPq).

\section*{Appendix}

Consider a $(M+1)-$dimensional orthogonal matrix $O$, and let $S$ be its $N$-dimensional
upper-left square corner. It was shown in \cite{khoruz} that, when $O$ is uniformly
distributed in the orthogonal group $\mathcal{O}(M+1)$ with Haar measure, $S$ becomes a
random matrix whose distribution is given by $P(S)=\frac{1}{\Z_T}\det(1-SS^T)^{M/2-N}$,
where the normalization constant is $\Z_T=\int dS \det(1-SS^T)^{M/2-N}.$

Suppose we wish to compute the average value of $\prod_{k=1}^{2n}O_{a_kb_k}$, with $1\leq
a_k,b_k\leq N$. Since all the elements belong to $S$, it is obvious that \be\label{A2O}
\left\langle
\prod_{k=1}^{2n}O_{a_kb_k}\right\rangle_{\mathcal{O}(M+1)}=\int\frac{dS}{\mathcal{Z}_T}\det(1-SS^T)^{M/2-N}
\prod_{k=1}^{2n}S_{a_kb_k}.\ee This is an exact relation between the statistics
properties of orthogonal matrices and those of general real matrices.

Introduce now the different normalization constant $\Z=\int dS e^{-\frac{M}{2}{\rm
Tr}SS^T}$, which is the one required in the use of Wick's rule. As discussed in the text,
we wish to compute the limit as $N\to 0$ of Eq.(\ref{A2O}), in the form \be \lim_{N\to
0}\frac{\Z}{\Z_T}\int\frac{dS}{\Z}\det(1-SS^T)^{M/2-N} \prod_{k=1}^{2n}S_{a_kb_k}.\ee It
is not difficult to show that $\lim_{N\to 0}\frac{\Z}{\Z_T}=1,$ and we therefore arrive
at Eq.(\ref{limit}).

\end{document}